\documentclass[aps,floats,floatfix,amssymb,amsmath,prb,twocolumn,superscriptaddress,nolongbibliography]{revtex4-2}
\usepackage[english]{babel}
\usepackage{amsmath}
\usepackage{bm}
\usepackage{graphicx,bbm} 
\usepackage{times}
\usepackage{epsfig}
\usepackage{float}
\usepackage{soul}
\usepackage[normalem]{ulem}
\usepackage{braket}
\usepackage[utf8]{inputenc}
\usepackage[colorlinks,linkcolor=blue,citecolor=blue,urlcolor=blue]{hyperref}
\usepackage{color}
\usepackage{latexsym}
\usepackage{ulem}
\definecolor{nred} {RGB}{224,0,0}
\definecolor{nblue} {RGB}{28,130,185}
\definecolor{pgreen}{RGB}{78,138,21}
\definecolor{norange}{RGB}{230,120,20}

\usepackage{amsmath}
\newcommand{\e}{\mathrm{e}}
\newcommand{\ii}{\mathrm{i}}
\newcommand{\dd}{\mathop{}\!\mathrm{d}}

\begin{document}


\title{Electronic Green's function across the pseudogap to stripe transition in the $t$-$t'$-$J$ model}
\author{Martin Ulaga}
\email{martinu@pks.mpg.de}
\affiliation{\it Max Planck Institute for the Physics of Complex Systems, Dresden, Germany}
\author{Aritra Sinha}
\affiliation{\it Max Planck Institute for the Physics of Complex Systems, Dresden, Germany}
\author{Alexander Wietek}
\email{awietek@pks.mpg.de}
\affiliation{\it Max Planck Institute for the Physics of Complex Systems, Dresden, Germany}

\begin{abstract}
Superconducting domes in strongly correlated electronic systems are often accompanied by charge density waves and peculiar features in the electronic structure. The appearance of a pseudogap, in particular, and its relation to charge density waves remains insufficiently understood. Here, we investigate the electronic Green's function of the underdoped $t$-$t'$-$J$ model of the cuprate superconductors using tensor network algorithms for finite temperature dynamics on cylinders of width $4$. We find the prominent momentum differentiation, the hallmark of the pseudogap, to be strongly dependent on $t'$, which develops into a momentum-dependent opening of a gap upon decreasing temperature, consistent with the formation of a Fermi arc at intermediate temperatures. The nodal gap around ${\bf k}=(\pi/2,\pi/2)$ closes and fills with increasing temperature as coherent stripe order melts into a regime of fluctuating charge clusters.
\end{abstract}

\maketitle


\textit{Introduction}---The pseudogap phenomenon, characterized by a partial suppression of electronic states near the chemical potential, and its relation to unconventional superconductivity remain a foremost challenge in contemporary condensed matter physics~\cite{timusk1999reppp,norman2005pseudogap}. Particularly intriguing is its relation to stripe order---an intertwined modulation of spin and density~\cite{kivelson03}---which may underlie the pseudogap phenomenon and compete with superconductivity~\cite{tranquada2020,devereaux2025}. Recent theoretical work has made this point explicit: diagrammatic Monte Carlo simulations of the Hubbard model~\cite{simkovic2024science} identified momentum differentiation at finite temperature, which extrapolated directly to ground state stripe order~\cite{xu2022prr}. 

Experimentally, the pseudogap has been extensively studied by angle-resolved photoemission spectroscopy (ARPES)~\cite{damascelli2003rmp,vishik2018rpp}, which has revealed the existence of several energy scales (single particle gaps) inside the pseudogap exhibiting pronounced anisotropies. Conversely, complementary pictures arising from scanning tunneling spectroscopy~\cite{schmidt2011njp} have highlighted the role of spatial inhomogeneity and electronic asymmetry. Together, these observations point to a complex phenomenology that challenges a description of pseudogap phenomena solely in terms of pre-formed pairs~\cite{fradkin2015rmp}.

From the theoretical side, the square lattice $t$-$J$ model has been widely used to approach the phenomenology of high-temperature superconducting cuprates, although the necessity to include extensions remains unsettled~\cite{ogata2008}. Early developments include exact diagonalization (ED) studies, which elucidated the role of further neighbour hopping terms for ARPES spectra~\cite{dagotto90,jaklic97,tohyama04,zemljic07,zemljic08}, with subsequent work including cluster perturbation theory \cite{zacher00,kohno15} and quantum Monte Carlo \cite{brunner00, 
mishchenko01}. Tensor network studies, beginning with DMRG evidence for stripes~\cite{white1998prl_stripes,white1999prb_ttprimej}, have played a central role in resolving the competing or coexisting ground-state orders, e.g., using algorithms based on matrix product states (MPS)~\cite{jiang18,jiang2021pnas,lu2024prl} or infinite projected entangled pair states~\cite{corboz2011prb,corboz2014prl}. More recent work has extended these approaches to target dynamical response functions~\cite{tohyama18} and finite-temperature properties 
\cite{qu24,guthardt2025prb,zhang2026prb}, where Ref.~\onlinecite{guthardt2025prb} discussed properties of the single-hole spectral functions on width-$4$ cylinders.

In this work, we connect ARPES signatures directly to stripe order by examining the evolution of the single-particle spectral function in a minimal $t-J$ model using finite-temperature matrix product states---specifically, minimally entangled typical thermal states (METTS)~\cite{white2009,stoudenmire2010,wietek2021prx}. The method is naturally suited for spatially inhomogeneous systems, since it operates in real space and does not assume translation invariance. It has recently been used to identify charge clustering in strongly correlated systems~\cite{sinha2025,sinha2026}. The significant development in our approach is to evaluate the dynamical correlation function on a contour parametrized by complex time, which allows us to probe the whole frequency dependence of the ARPES signal in a controlled way, avoiding the usual entanglement bottleneck of MPS. We can thus directly track the emergence of ground-state stripe order and its imprint on the single-particle spectrum at low temperature, while recovering the characteristic momentum differentiation at intermediate temperature. In particular, we observe a characteristic gap opening around ${\bf k}=(\pi/2,\pi/2)$---often referred to as the nodal region where the electronic gap should close for a $d$-wave superconductor---as temperature is lowered towards the stripe regime. 


\begin{figure*}[t]
    \centering
    \includegraphics[width=0.49\linewidth]{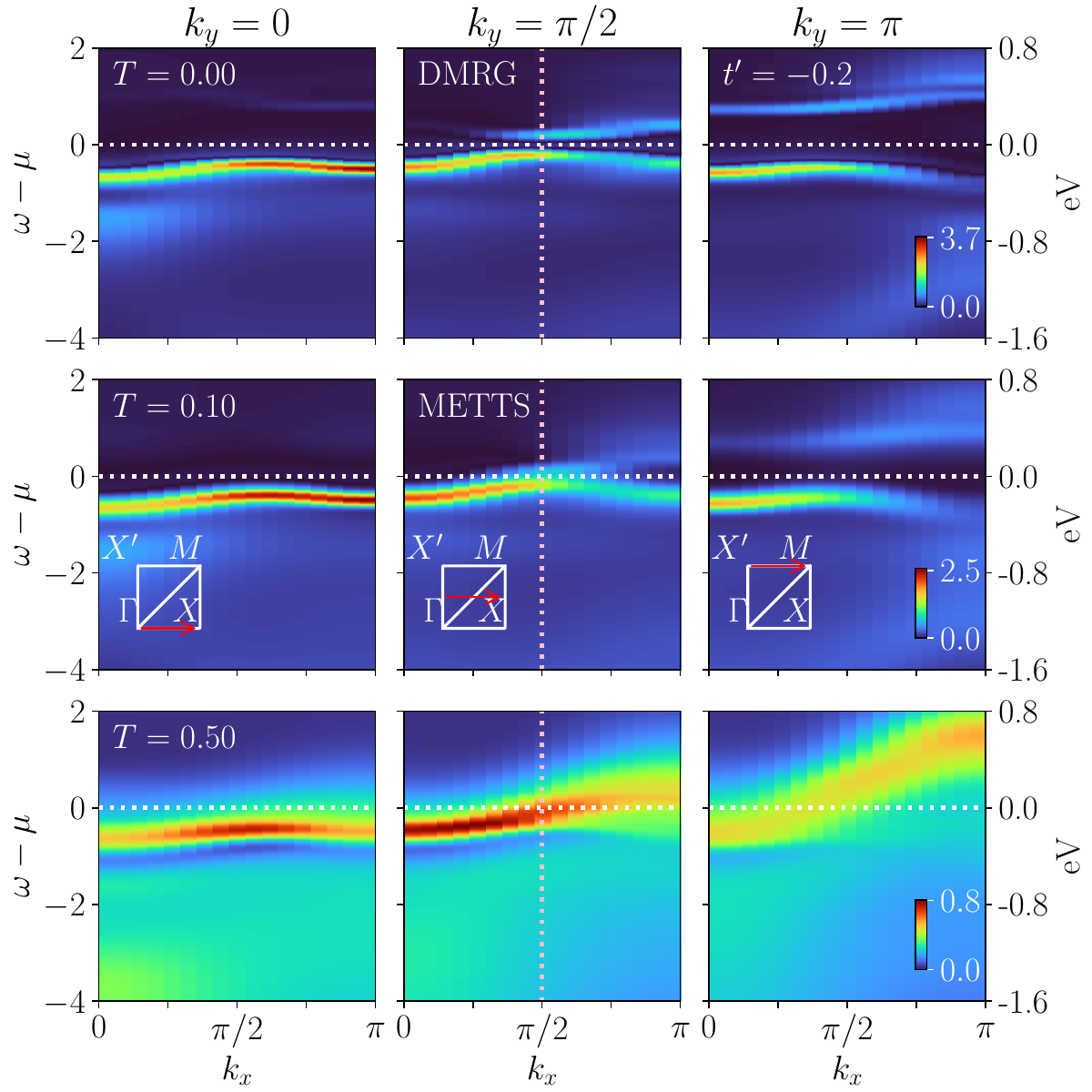}
    \includegraphics[width=0.49\linewidth]{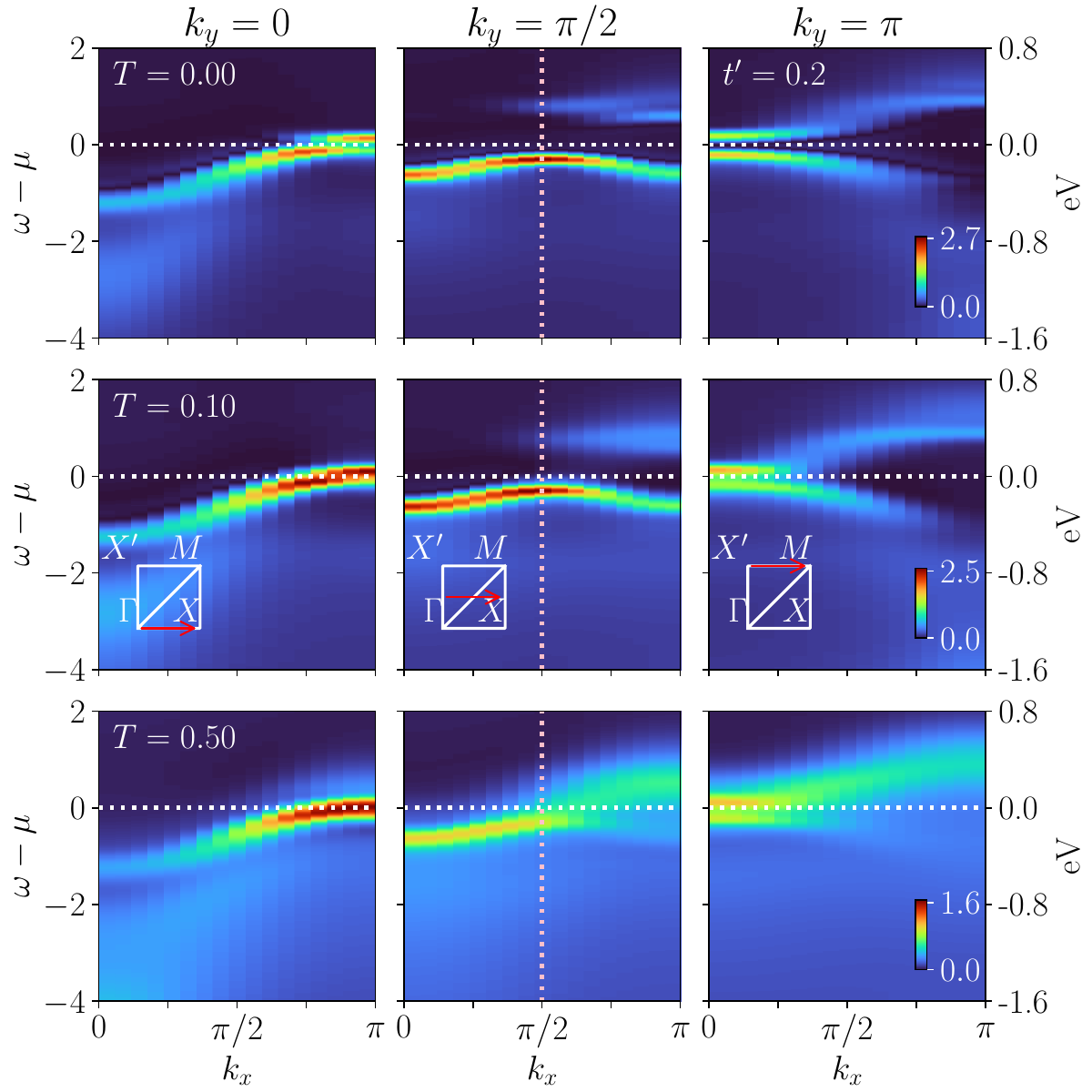}
    \caption{The single-particle spectral function $A({\bf k},\omega)$ for three momentum cuts with $k_y=0, \pi/2 ,\pi$ at $T=0$ (DMRG), $T=0.1$, $T=0.5$ at doping $p=1/16$ and $t'=-0.2$ (left, relevant for hole-doped cuprates), or $t'=0.2$ (right). The dynamical METTS were taken from up to $R=20$ independent Markov chains running up to 150 steps deep and time-evolved to $\Omega_t=50$ using a complex angle $\theta=\pi/16$, cf. Appendix~\ref{app:tdvp}. The energies $\omega$ are taken with respect to the chemical potential computed as in Eq.~\ref{eq:mudef}. For comparison to ARPES results, we employ a common estimate $t=0.4\,$eV~\cite{ogata2008}. For both $t'=0.2$ and $t'=-0.2$, we observe two well-developed bands separated by a gap around the chemical potential, which merge in the nodal region around ${\bf k}_N = (\pi/2, \pi/2)$ upon increasing temperature. The lower band crosses the chemical potential only for $t'=0.2$ but not for $t'=-0.2$.}
    \label{fig:aw_tp}
\end{figure*}

\textit{Model and observables}---We study the $t$-$t'$-$J$ model \cite{spalek77}, given by
\begin{multline}
    H=-t\sum_{\langle ij\rangle\sigma}(c^{\dagger}_{i\sigma} c_{j\sigma} +\textrm{h.c.})-t'\sum_{\langle\langle ij\rangle\rangle\sigma}(c^{\dagger}_{i\sigma} c_{j\sigma} +\textrm{h.c.}) \\
    +J\sum_{\langle ij\rangle}({\bf S}_i\cdot {\bf S}_j-\frac{1}{4}n_in_j)
    \label{eq:ham}
\end{multline}
\noindent on a finite cylinder of width $4$. Above, $c_i$ ($c_i^{\dagger}$) are fermion annihilation (creation) operators with the restriction on double occupancy implied, and $n_i=\sum_{\sigma}c^{\dagger}_{i\sigma}c_{i\sigma}$. We set $t=1$ as the unit of energy and fix $J=0.4$, corresponding to a strong coupling ($U/t=10$) Hubbard model, and consider $t'=\pm 0.2$. This choice of $t'$ is often used to mimic the particle-hole asymmetry between electron-doped ($t'=0.2$) and hole-doped ($t'=-0.2$) cuprate phenomenology~\cite{ogata2008}. The chosen parameter regime on the finite cylinder is notable because it has been observed to host superconducting and striped ground states \cite{corboz2011prb,jiang18,jiang2021pnas, wietek2022prl, lu2024prl}, but also a tendency to phase separation \cite{emery90prl} or clustering at finite temperature \cite{sinha2025}.

\noindent We investigate the real-time retarded Green's function~\cite{mahan1981},
\begin{equation}
    G^{\textrm{R}}({\bf k}, t - t') = -\ii \theta(t-t')\sum_{\sigma=\uparrow,\downarrow}\braket{\{c_{{\bf k}\sigma}(t),\, c^\dagger_{{\bf k}\sigma}(t')\}}_\beta,
\end{equation}
where $\braket{\mathcal{O}}_\beta = \text{Tr}[\e^{-\beta H} \,\mathcal{O}] / \mathcal{Z}$ denotes a thermal expectation value of an operator $\mathcal{O}$ at inverse temperature $\beta = 1/k_{\text{B}}T$ ($k_{\text{B}}=1$ henceforth). $\theta(t)$ denotes the step function, $\{A,B\}=AB + BA$, and the fermionic (annihilation) operators at momentum $\bf k$ are given as
\begin{equation}
    c_{{\bf k}\sigma}=\frac{1}{\sqrt{N_s}}\sum_j\e^{\ii{{\bf k}\cdot {\bf r}}_j}\,c_{j\sigma},
\end{equation}
where $\sigma=\uparrow,\downarrow$, ${\bf r}_j$ denotes the $j$-th lattice position and $N_s$ the number of lattice sites. The frequency-dependent Green's function is given by
\begin{equation}
    G^{\textrm{R}}({\bf k}, \omega) = \int_{-\infty}^\infty\dd t\, \e^{\ii \omega t}\,G^{\textrm{R}}({\bf k}, t)
\end{equation}
and the spectral function by $A({\bf k}, \omega) = -(1/\pi)\, \textrm{Im} \,G^{\textrm{R}}({\bf k}, \omega)$. To perform spatial and temporal Fourier transforms, we apply windowing functions, cf. Appendix~\ref{app:tdvp}. The spectral function is related to the experimentally measured ARPES intensity via
\begin{equation}
    I({\bf k}, \omega) \propto f(\omega) A({\bf k}, \omega),
\end{equation}
where $f(\omega) = (\e^{\beta(\omega - \mu)}+1)^{-1}$ is the Fermi function~\cite{damascelli2003rmp}. The chemical potential $\mu$ is therefore crucial for the comparison to ARPES. As our METTS and DMRG simulations are performed in the canonical ensemble, $\mu$ is not a parameter, but instead determined by a Legendre transformation, 
\begin{equation}
\label{eq:mudef}
    \mu(N) = \{ \nu\; | \; \mathop{\textrm{argmin}}_{M} (F(M) - \nu M)=N\},
\end{equation}
where $F(M)$ denotes the free energy with $M$ particles. This construction yields an interval, whose midpoint we, henceforth, refer to as simply $\mu(N)$. Determining $\mu$ via a Legendre transformation instead of a simple derivative is required, as the ground state energy $E_0$ is a non-convex function of $N$, due to the pair-binding effects, cf. Appendix \ref{app:chemical_potential} for details.


\begin{figure*}[t]
    \centering
\includegraphics[width=0.95\linewidth]{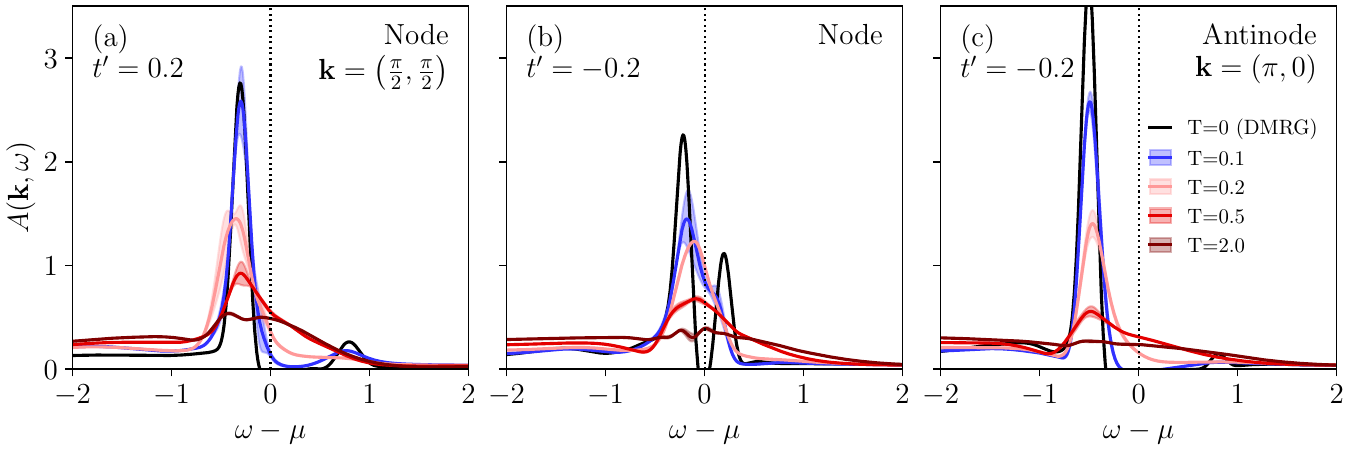}
    \caption{Spectral function $A({\bf k}, \omega)$ of the $t$-$t'$-$J$ model on the $16\times 4$ cylinder for $J=0.4$ as a function of temperature obtained using dynamical METTS at the node (${\bf k}=(\pi/2,\pi/2)$) for $t'=\pm0.2$ (a,b) and the antinode (${\bf k}=(\pi,0)$) for $t'=-0.2$ (c). In all cases, $\theta=\pi/16$. The vertical dashed line indicates the value of the chemical potential $\mu$. In each case, we observe two peaks that gradually merge into a single peak upon increasing temperature, filling the gap between the two distinct bands.}
    \label{fig:cuts_temp}
\end{figure*}

\textit{Dynamical METTS}---We employ the recently developed dynamical METTS method for the evaluation of dynamical correlation functions at finite temperature~\cite{wang2025prb,wang2025prl}. Within the METTS method, a thermal observable is evaluated using
\begin{equation}
    \braket{\mathcal{O}}_\beta=\frac{1}{\mathcal{Z}}\textrm{Tr}(\e^{-\beta H} \mathcal{O}) = \frac{1}{\mathcal{Z}}\sum_i p_i \braket{\psi_i|\mathcal{O}|\psi_i},
\end{equation}
where $p_i = \braket{\sigma_i|\e^{-\beta H} | \sigma_i} \geq 0$ with
\begin{equation}
    \ket{\psi_i} = \frac{1}{\sqrt{p_i}} \e^{-\beta H/2} \ket{\sigma_i},
\end{equation}
where $\ket{\sigma_i} = \ket{\sigma_i^1}\cdots\ket{\sigma_i^N}$ denotes the basis of product states. The states $\ket{\psi_i}$ are referred to as METTS snapshots, and are sampled with probabilities $p_i/\mathcal{Z}$ according to the METTS algorithm~\cite{white2009,stoudenmire2010}. The METTS snapshots $\ket{\psi_i}$ obtained from imaginary-time evolution of product states are computed using matrix-product state (MPS) techniques, see Ref.~\cite{wietek2021prx} for details on the simulation. Statistical averages for expectation values are obtained by random averaging over $R$ different snapshots, $\braket{\mathcal{O}}_\beta \approx R^{-1}\sum_{i=1}^R \braket{\psi_i|\mathcal{O}|\psi_i}$.
For dynamical observables of the form $\mathcal{O} = A(t)B$  we introduce the states
\begin{align}
&|v_i(t)\rangle = \e^{-\ii Ht} B |\psi_i \rangle \label{eq:auxstatevi},\\
&|w_i(t)\rangle = \e^{-\ii Ht} |\psi_i \rangle, \label{eq:auxstatewi}
\end{align}
which are explicitly computed using MPS time-evolution techniques up to a final time $\Omega$, and evaluate
\begin{equation}
    \braket{\psi_i|A(t)B|\psi_i} = \braket{w_i(t)|A|v_i(t)}.
\end{equation}
To alleviate the computational burden of entanglement growth upon real-time evolution, we compute complex-time \textit{Hermitian} correlators \cite{wang2025prb,grundner2023},
\begin{equation}
    \braket{A(z)B} = \braket{\e^{\ii H \bar{z}} A \e^{-\ii H z}B},
\end{equation}
with complex time $z=t-\ii \tau$ chosen on a complex contour with small complex angle $\theta$, such that in the limit $\theta\rightarrow0$, the real-time limit is retrieved. A detailed discussion regarding convergence with MPS bond dimension, cylinder length $L$, and complex angle $\theta$ is found in Appendix~\ref{app:tdvp}.


\textit{Electronic structure}---We first present results for $A({\bf k},\omega)$ for $T=0, 0.1, 0.5$ in Fig.~\ref{fig:aw_tp} for the case of $t'=0.2$ and $t'=-0.2$ at doping $p=1/16$. Several spectral features are present already at $T=0$. First, a bright and rather sharp dispersive band below $\mu$ with bandwidth $\approx 1.3t$ for $t'=0.2$ (further flattened to $\approx 0.6t$ for $t'=-0.2$), which is observed also for $t'=0$ and $p=0$, cf. App.~\ref{app:heatmaps}, indicating its origin in hole excitations. The breakdown of the spectral function into particle and hole contributions is presented in App.~\ref{app:tdvp}. In all cases, this bandwidth is significantly smaller than the bandwidth of a noninteracting model on the same lattice, $8t$. Since magnetic interactions strongly renormalize this band, we interpret this feature as a magnetic polaron, i.e., a hole dressed in deformations of the antiferromagnetic background~\cite{martinez1991}. Interestingly, the polaron band has a maximum near the node. Secondly, incoherent signatures are present at ${\bf k}=(0,0)$ and $(\pi,\pi)$. Thirdly, there is a rather nondispersive (in the $x$ direction) gapped particle excitation at $k_y=\pi/2$. The effect of the geometry manifests in $A({\bf k},\omega)$ as an only approximate equivalence of the antinodal points ${\bf k}_A=(\pi,0)$ and ${\bf k}_{A'}=(0,\pi)$ (i.e., the cylinder is not $C_4$-symmetric). This distinction is particularly present at $T=0$. 

This general picture at $T=0$ appears quite universal across various $t'$ (cf. App.~\ref{app:heatmaps}), which has also been observed not to disturb CDW formation in the ground state, although the CDW amplitudes are typically influenced by $t'$ \cite{jiang2021pnas}. While the nodal gap $\Delta_N$ appears across the considered $t'$, albeit with varying magnitude, both the hole and particle excitation components at the antinode show a more intricate dependency. In particular, for $t'=0.2$ the antinodal gap $\Delta_{A'}$ is not distinguishable from the connected band, but becomes very pronounced for $t'=-0.2$.

Increasing $T$ has three effects: (a) the spectral weight at the node ${\bf k}_N=(\pi/2,\pi/2)$ picks up as the nodal gap $\Delta_N$ melts, (b) the spectral features become universally broader, which results in (c) the polaron band in the $X'-M$ direction forming a ``waterfall'' feature with the incoherent particle excitations, eventually merging into a single band~\cite{zemljic08}. 


Next, we show $A({\bf k},\omega)$ for selected ${\bf k}_{N/A}$ in Fig.~\ref{fig:cuts_temp} for several values of $T$. As the temperature is increased, the gap between polaron states below the chemical potential and particle excitations rapidly melts, with the ``activation'' being mainly dictated by the size of the underlying bandstructure gaps. For example, the case of $t'=0.2$, Fig.~\ref{fig:cuts_temp}(a), shows a $\Delta_N\approx t$, but the proximity of the polaron to the chemical potential results in the gap filling already for $T\approx 0.1$. The converse is observed for $t'=-0.2$, Fig.~\ref{fig:cuts_temp}(b-c), where the dominant gap is at the antinode, $\Delta_A\approx 1.5$, and remains open until a higher $T\approx 0.2-0.5$ since the polaron band is pushed farther below the chemical potential. The activation of $\Delta_A$ thus happens only after $\Delta_N$ has already melted.

\begin{figure}[t]
    \centering
    \includegraphics[width=0.95\linewidth]{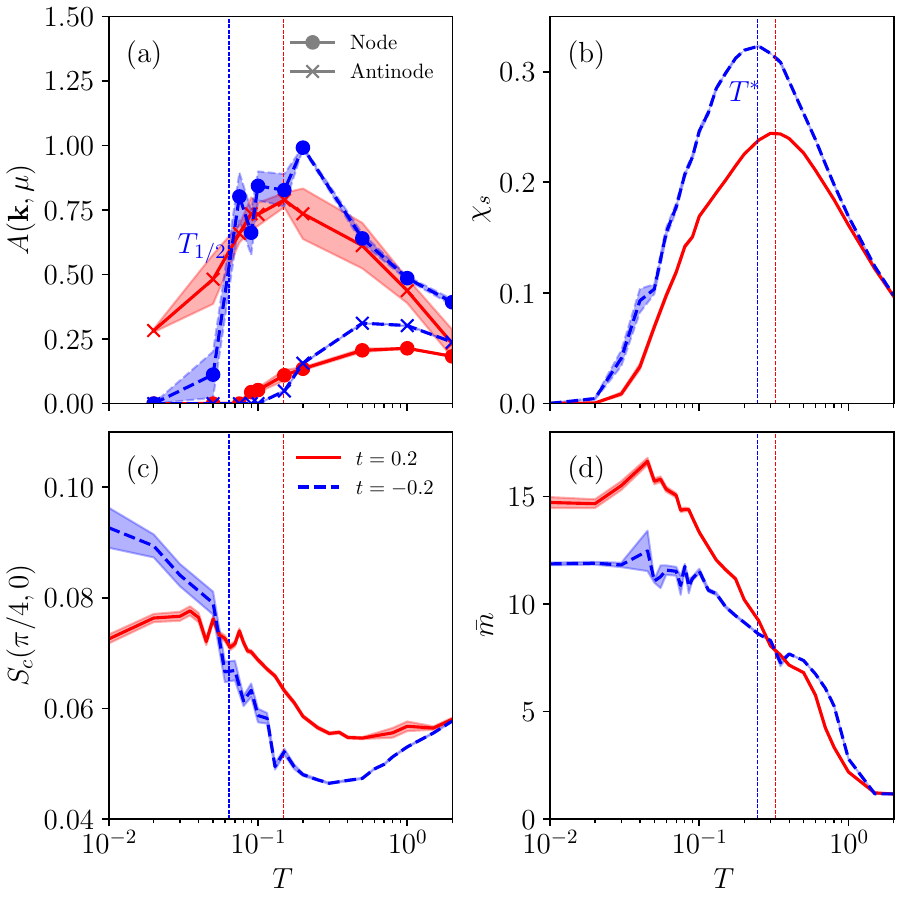}
    \caption{Temperature dependence of the (a) spectral weight at the chemical potential at the node (${\bf k}_N = (\pi/2, \pi/2)$) and the antinode (${\bf k}_A = (\pi, 0)$) (b) uniform magnetic susceptibility $\chi_s$ (c) charge structure factors at the CDW ordering wave-vector ${\bf{k}}_s= (\pi/4, 0)$ (d) mean charge cluster size. The suppression of nodal spectral weight at the chemical potential (vertical dashed lines, obtained as the half of the maximum intensity) occurs at the onset of CDW formation, denoted by $T_{1/2}$, e.g., $T_{1/2}\approx0.12$ for $t'=0.2$. $T_{1/2}$ coincides with the rapid onset of CDW order, quantified by $S_c({\bf k}_s)$ in panel (c). The maximum of $\chi_s$ defines the pseudogap onset temperature $T^*$, e.g., for $t'=0.2$, $T^*\approx 0.32$.
    }
    \label{fig:3}
\end{figure}

\textit{Thermodynamics and the real-space picture}---To connect the phenomenology of the spectral function to thermodynamic instabilities, we show the spectral intensity at the chemical potential $A({\bf k},\mu)$ as a function of $T$ alongside the charge structure factor,
\begin{equation}
S_c({\bf k})=\frac{1}{N_s}\sum_{ij}e^{-\ii \mathbf{k}\cdot (\mathbf{r}_i-\mathbf{r}_j)}\langle(n_i-\langle n\rangle)(n_j-\langle n\rangle)\rangle,
\end{equation}
\noindent and the uniform magnetic susceptibility, 
\begin{equation}
\chi_s(T)=\frac{\dd M}{\dd H} =\frac{\langle M^2\rangle - \langle M\rangle^2}{N_sT},
\end{equation}
where $M=\sum_i S^z_i$, in Fig.~\ref{fig:3}. One sees from Fig.~\ref{fig:3}(a,c) that nodal spectral weights drop as $S_c({\bf k}_s)$, where ${\bf k}_s=(\pi/4,0)$ is the CDW wave-vector for the particular cylinder geometry, rises at low $T$. To quantify this observation, we define $T_{1/2}$ as the temperature at which $A({\bf k}_N,\mu)$ falls to half its maximum value, and analogously, the CDW formation temperature $T_{1/2}^{(c)}$ from $S_c({\bf k}_s)$. We find, for $t'=0.2$, that $T_{1/2}\approx T_{1/2}^{(c)}\approx 0.12$. Likewise for $t'=-0.2$, $T_{1/2}\approx T_{1/2}^{(c)}\approx0.06$. The pseudogap crossover temperature $T^*$ is obtained from the maximum of $\chi_s$, see Fig.~\ref{fig:3}(b), $T^*\approx 0.3$. Below, $\chi_s$ is consistent with a spin gap. The rise in nodal spectral weight is thus found to be consistent with the thermal melting of the stripe-induced gap, seen in Fig.~\ref{fig:aw_tp}. 

As $T$ increases, the intensity at the node rapidly fills up on crossing the ``threshold'' $T\sim 0.1$, where the stripe exhibits melting and crosses over to a regime of ``forestalled phase separation'' \cite{sinha2025}. In particular, once stripe order melts, the doped charge redistributes into fluctuating hole-rich clusters of varying size, producing strong low-momentum charge inhomogeneity that can scatter nodal quasiparticles efficiently.


To make the real-space origin of this crossover explicit, we analyze charge clustering in individual METTS snapshots. For each snapshot, we first compute the local hole density
$n_h^{(s)}(\mathbf r)=1-\langle\psi_s|n(\mathbf r)|\psi_s\rangle$.
We then mark the sites whose hole density is larger than an adaptive threshold based on the mean and variance of $n_h^{(s)}(\mathbf r)$ (see Refs.~\cite{sinha2025, sinha2026} or Appendix~\ref{app:clustering} for details). Neighboring marked sites are grouped together to form a cluster $\mathcal C$. For each cluster, we record its size $m_{\mathcal C}$, i.e., how many sites it occupies, and its hole mass $M_{\mathcal C}=\sum_{\mathbf r\in\mathcal C} n_h^{(s)}(\mathbf r)$, i.e., how much doped charge it carries. We then construct the density-weighted cluster-size distribution $p_m$, which tells us what fraction of the total doped charge sits in a cluster of size $m$. Fig.~\ref{fig:3}(d) shows the mean cluster size $\bar m=\sum_m m\,p_m$ as a function of temperature. High values of $\bar m$ at temperatures above the onset of stripe order signal strong hole clustering.

\begin{figure}[t]
    \centering
    \includegraphics[width=0.95\linewidth]{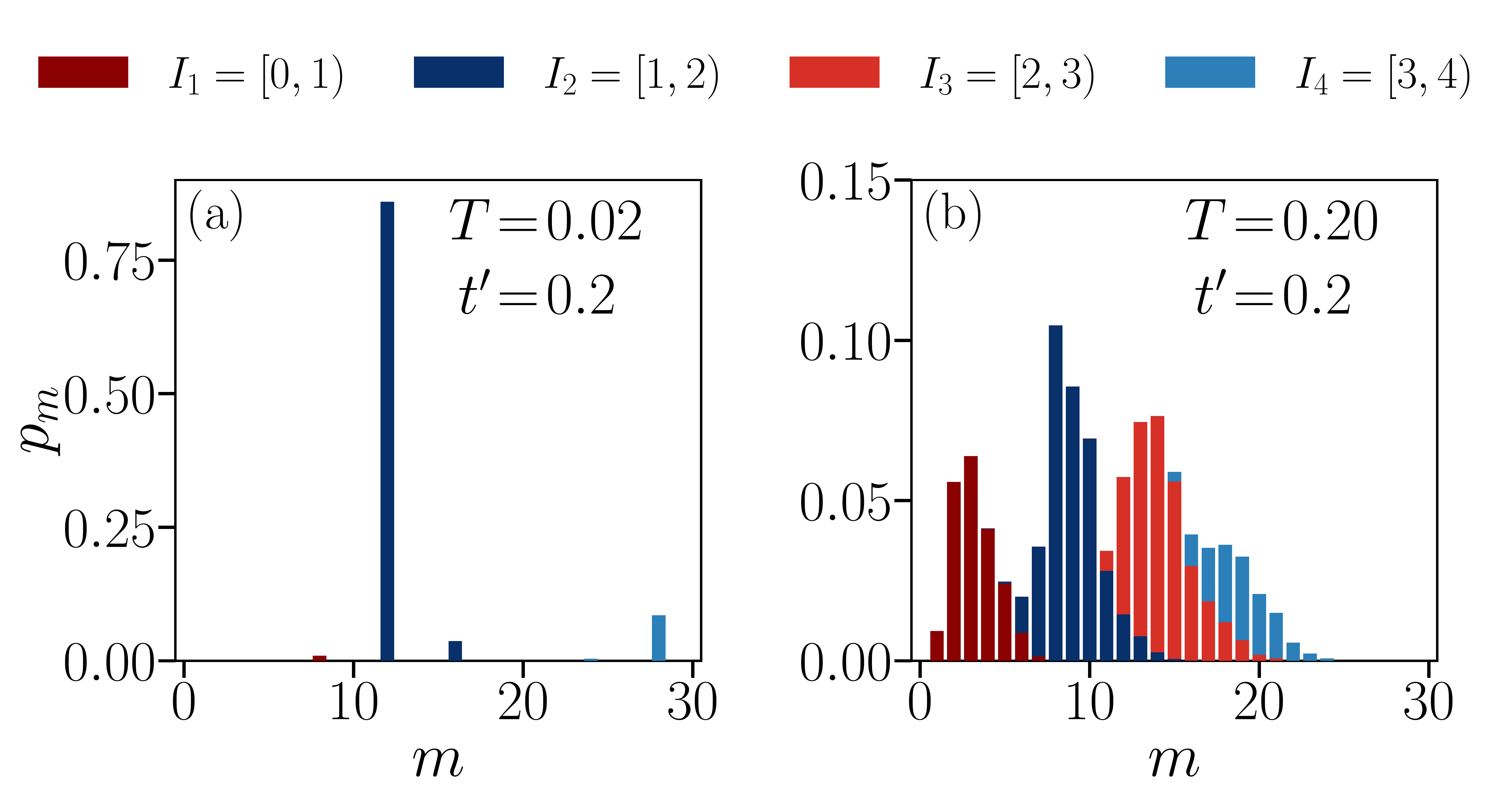}
    \caption{
    Density-weighted cluster-size distributions $p_m$ for the $16\times4$ cylinder at $p=1/16$ and $t'/t=0.2$, extracted from METTS snapshots. The total height at fixed $m$ gives the fraction of doped charge residing in hole-rich connected clusters of size $m$, while the stacked colors resolve this weight by the cluster hole mass in the intervals $I_1=[0,1)$, $I_2=[1,2)$, $I_3=[2,3)$, and $I_4=[3,4)$. (a) At $T=0.02$, the weight is sharply concentrated near a characteristic size, consistent with charge density waves. (b) At $T=0.20$, the weight is spread over many sizes, indicating a fluctuating mesoscopic clustered regime after stripe melting.}
    \label{fig:clustering}
\end{figure}

In Fig.~\ref{fig:clustering}, we show the probability $p_{m}$ of a cluster of size $m$. The stacked colors resolve how many holes these clusters of a certain size contain, using the mass intervals $I_1=[0,1)$, $I_2=[1,2)$, $I_3=[2,3)$, and $I_4=[3,4)$. Thus, the horizontal axis measures spatial extent, the bar height measures statistical importance, and the colors indicate hole charge content.

Figure~\ref{fig:clustering} shows that the melting of the stripe does not produce a featureless state. At $T=0.20$, the doped charge is distributed over clusters of many different sizes. Upon cooling to $T=0.02$, the weight collapses onto a characteristic size $m\approx 12$, corresponding to the typical stripe cluster size on this geometry~\cite{sinha2025}. The nodal gap therefore fills not simply because of generic thermal broadening, but because coherent stripe order gives way to a distinct intermediate-temperature regime with strong real-space charge clustering. This is the sense in which the system exhibits forestalled phase separation: the holes strongly cluster, but at intermediate temperature, they remain distributed among finite clusters rather than collapsing into a single macroscopic cluster.

\textit{Discussion}---Using the recently developed dynamical METTS~\cite{wang2025prb,wang2025prl}, we have followed signatures of the pseudogap phenomenon down in temperature to the striped ground state in the $t$--$t'$--$J$ model.
At doping $p=1/16$, associated with the underdoped regime in cuprates, the ground state on the studied $W=4$ cylinder exhibits a CDW, and the spectral function exhibits two band-like features separated by a gap, whose size is strongly momentum dependent.  An important parameter in the $t$-$J$ model is next-neighbour hopping $t'$, which controls the electron- or hole-like shape of the noninteracting Fermi surface~\cite{tohyama04,ogata2008}. Usually, $t'$ in the range $(-0.1,-0.3)$ is taken to model hole-doped cuprates~\cite{ogata2008}. However, recent studies often suggest that pairing is enhanced only for $t'>0$~\cite{jiang2021pnas,chen2025,zhang2026prb,li2026fluctuatingpairdensitywave}, although the boundary conditions may also play an important role~\cite{jiang18}. On the finite cylinder studied here, we observe that $t'$ strongly influences the renormalized polaron band and controls the gaps at the antinode.

A long-standing discussion in the physics of cuprate superconductors is the occurrence of Fermi arcs or Fermi pockets in the pseudogap regime~\cite{Sebastian2012}. While current ARPES experiments measure Green's functions consistent with Fermi arcs~\cite{Yoshida2012,vishik2018rpp}, other observations, such as quantum oscillations~\cite{LeBoeuf2007,DoironLeyraud2007,Bangura2008,Yelland2008} or the Yamaji effect from recent magnetoresistivity measurements in HgBa$_2$CuO$_{4+\delta}$~\cite{Chan2025,Zhao2025} exhibit signatures of Fermi pockets~\cite{Yang2010}. The question thus remains whether the $t$-$J$ model captures either of the two scenarios. While for $t'=-0.2$ at zero temperature, we observe a full gap throughout the Brillouin zone, this gap has been shown to close in the nodal regime first upon melting the CDW in the pseudogap regime, while a gap at the antinode remains open until higher temperatures. The Green's function only crosses the chemical potential $\mu$ once at $T=0.1$ for the momentum cut from $(0, \pi/2)-(\pi, \pi/2)$, which is consistent with the Fermi arc scenario. However, we note that the polaron band exhibits a local maximum at the nodal point. Hence, if by some mechanism the chemical potential were to be lowered, such that it crosses the Green's function twice, such an observation could be indicative of the Fermi pocket scenario. It remains conceivable that a more elaborate model, such as the three-band Emery model~\cite{Emery1987}, could indeed exhibit Fermi pockets. Finally, we note that our findings are also consistent with growing Fermi arcs upon increasing temperature,
as observed for example in Bi$_2$Sr$_2$CaCu$_2$O$_{8+\delta}$ ~\cite{Kanigel2006}, since we find that our gap first closes at the node, and only at higher temperatures is the antinodal gap closed. Thus, the formation of Fermi arcs within the $t$-$t'$-$J$ model is a genuinely finite-temperature phenomenon and not a property of the ground state. 

The nodal gap $\Delta_N$ appears in the renormalized polaron band independently of $t'$, therefore offering an interpretation of the gap in terms of CDW formation. This is consistent with real-space DMFT studies of the Hubbard model~\cite{fleck01prb,raczkowski2010prb}. Reproducing this observation in ARPES has, however, proven challenging as inhomogeneous or fluctuating stripes can obscure spectral features associated with a nodal gap~\cite{kivelson03}. Static stripes do not suffer from this problem and have been associated with such signatures in La$_{2-x}$Sr$_x$CuO$_4$ (LSCO)~\cite{ino2000prb,razzoli2013prl, matt2015}. Interestingly, CDW order has been recently found to coincide with the pseudogap upon the suppression of superconductivity in Eu-LSCO~\cite{missiaen2025prx}. Nevertheless, alternative routes for probing the interplay of momentum differentiation and CDW formation are also desired, such as cold atom simulators, which are now reaching the relevant temperature scale~\cite{chalopin2026pnas}. For example, a recent lattice spectroscopy study combined with susceptibility measurements supports the connection between charge instability and partial Fermi surface depletion~\cite{kendrick2025}.

We also note a pronounced particle-hole asymmetry, i.e., $A({\bf k}, \omega-\mu)\neq A(-{\bf k},-(\omega-\mu))$, below $T^*\approx0.3$ where $\Delta_N$ appears for the case $t'=0.2$ (equivalently, $\Delta_A$ for $t'=-0.2$). In Ref.~\onlinecite{hashimoto2010nphys}, particle-hole asymmetry has been reported for Pb$_{0.55}$Bi$_{1.5}$Sr$_{1.6}$La$_{0.4}$CuO$_{6+\delta}$ whose origin is argued to lie within spatial symmetry breaking in the form of a CDW, interestingly, with short charge correlation lengths being preferred. Our observation that hole clustering is strongly enhanced between $T_{1/2}<T<T^*$ at the onset of momentum differentiation could thus offer support for the importance of spatial inhomogeneity in pseudogap phenomenology. 

On the technical side, the use of complex-time Hermitian correlation functions~\cite{wang2025prb} allowed us to reduce computational effort while retaining numerical accuracy. Complex-time correlation functions have only recently been demonstrated as a highly effective computational method for estimating Green's functions~\cite{Cao2024,grundner2023,wang2025prl}, as they naturally limit the entanglement of time-evolved matrix product states. Moreover, a novel approach has now demonstrated that the Nyquist-Shannon limit can be overcome by complex-time Krylov expansion~\cite{Paeckel2026}. As such, complex-time correlation approaches are now increasingly allowing the study of spectral properties of strongly correlated electron systems on larger systems with increased accuracy, especially at finite temperatures, where finite-size effects are diminished due to typically decreasing correlation lengths. We note that several recent works on wider cylinder geometries with similar parameters are nevertheless consistent with CDW ground states~\cite{jiang2021pnas,lu2024prl}. In this light, the spectral function's temperature evolution near the node is unlikely to be a quasi-1D artifact, but rather reflects intrinsic physics of the CDW-ordered state. Our findings are thus consistent with momentum differentiation as reported in a recent tensor network study~\cite{qu24}.


We have here focused on a strongly coupled $t-J$ model, so it would be exciting to extend the picture to the Hubbard model at intermediate to weak coupling. Recent work in this context has observed finite-temperature signatures of pseudogap in a region of the phase diagram connected to stripe formation~\cite{simkovic2024science,li2026fluctuatingpairdensitywave}. Our findings are compatible with this picture, but are crucially able to track the evolution of the pseudogap through the stripe formation temperature. We leave this matter for future work.

The pseudogap phenomenon extends beyond superconducting cuprates, e.g., it has also been observed in nickelates~\cite{uchida2011prl,zhao2021prl}, thin layers of conventional superconductors \cite{sacepe2010natcom}, ferromagnetic manganites~\cite{mannella2005nature}, and iridates~\cite{battisti2017natphys,hsu2024natphys}. Crucially, some of these systems do not superconduct, suggesting the pseudogap does not necessarily originate in pairing. Whether the mechanism driven by a CDW instability identified here applies more broadly remains an open question, but the ubiquity of pseudogap signatures alongside charge inhomogeneities suggests common underlying physics. 

\begin{acknowledgments} We thank J. Mravlje, R. {\v Z}itko, and T. Tohyama for insightful discussions. We further thank T. Tohyama and P. Prelov{\v s}ek for useful comments on an early version of this manuscript. A.W. acknowledges support by the German Research Foundation (DFG) through the Emmy Noether program (Grant No. 509755282). Funded by the European Union (ERC, MoNiKa, 101220368). Views and opinions expressed are however those of the author(s) only and do not necessarily reflect those of the European Union or the European Research Council. Neither the European Union nor the granting authority can be held responsible for them. A.S. acknowledges the Alexander von Humboldt Foundation for support under the Humboldt Research Fellowship.
\end{acknowledgments}

\bibliography{ref}

\newpage
\appendix

\renewcommand{\thefigure}{S\arabic{figure}}
\setcounter{figure}{0}

\section{Chemical potential}
\label{app:chemical_potential}
\begin{figure}
    \centering
    \includegraphics[width=\linewidth]{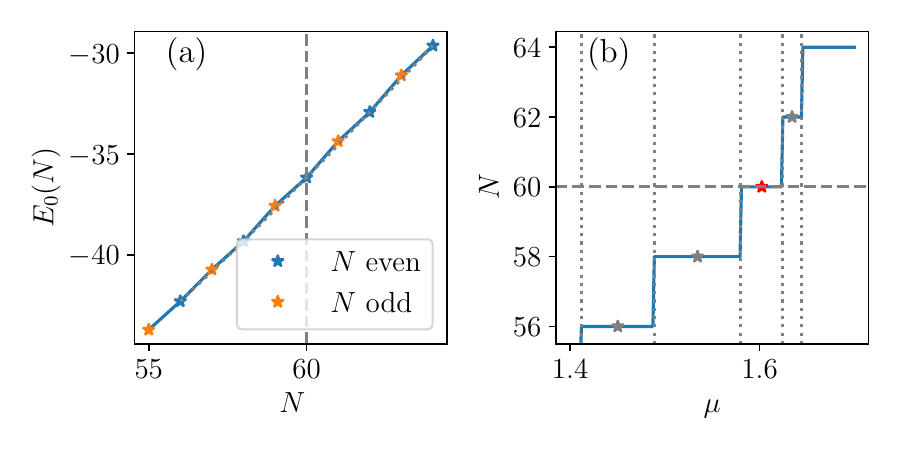}
    \caption{(a) Non-convex ground state energy of the     $t$-$t'$-$J$ model on the $16\times4$ cylinder for $t'/t=0$ and $J/t=0.4$. $E_0(N)$ lies above the line connecting $E_0(N-1)$ and $E_0(N+1)$ for $N$ odd. (b) particle number $N$ as a function of the chemical potential $\mu$ after the Legendre transformation in Eq.~\ref{eq:legendresup}.}
    \label{fig:chemical_potential}
\end{figure}

Determining the value of the chemical potential is crucial for interpreting the gap structure and the comparison to experimental ARPES data, due to the fact that processes above the chemical potential are suppressed by the Fermi factor in the ARPES intensity~\cite{damascelli2003rmp},
\begin{equation}
    I(\bf{k}, \omega) \propto f(\omega) A(\bf{k}, \omega).
\end{equation}
where $f(\omega) = (\e^{\beta(\omega - \mu)}+1)^{-1}$. Although our METTS and DMRG simulations are run in the canonical ensemble, a chemical potential can be defined by,
\begin{equation}
\label{eq:chempotentialderivative}
    \mu = \frac{\dd F}{\dd N},
\end{equation}
if the free energy $F$ were a convex function of $N$. However, as shown in Fig.~\ref{fig:chemical_potential}(a), the ground state energy $E_0(N)$ of the $t$-$t'$-$J$ model on the $16\times4$ cylinder for $t'/t=0$ and $J/t=0.4$ is not a convex function of the particle number $N$. Instead, we observe an even-odd effect where $E_0(N)$ lies above the line connecting $E_0(N-1)$ and $E_0(N+1)$ for $N$ odd. Physically, this even-odd effect is a pair-binding effect, where the energy difference of introducing two holes is less than twice that of introducing a single hole. This is typically measured by the pair binding energy,
\begin{equation}
    \Delta_p = E_0(N-2) - 2E_0(N-1) + E_0(N),
\end{equation}
which in this case is negative, $\Delta_p <0$, as previously reported in e.g.~\cite{white1997}.

Thus, the thermodynamically correct procedure of obtaining a conjugate variable is to perform a Legendre transformation~\cite{zia2008}. For this, one usually considers the conjugate free energy,
\begin{equation}
    \Omega(\mu) = \min\limits_N (F(N) - \mu N),
\end{equation}
and analogously,
\begin{equation}
\label{eq:legendresup}
    N(\mu) = \mathop{\textrm{argmin}}\limits_M (F(M) - \mu M).
\end{equation}
The function $N(\mu)$ for the $16\times4$ cylinder for $t'/t=0$ and $J/t=0.4$ is shown in Fig.~\ref{fig:chemical_potential}(b). Notice that the number of particles increases by $2$ instead of $1$, consistent with the non-convexity of $E_0$. To obtain the chemical potential $\mu(N)$, we would need to invert the function $N(\mu)$. However, since $N(\mu)$ is a step function, its inversion is multivalued, where for each value of $N$ we obtain an interval of $\mu$,
\begin{equation}
    \mu(N) = \{ \nu\; | \; \mathop{\textrm{argmin}}_{M} (F(M) - \nu M)=N\}.
\end{equation}
Thus, the chemical potential in a strict sense is not uniquely defined. To assign a single number for the chemical potential, we choose the midpoint of these intervals, as indicated by the stars in Fig.~\ref{fig:chemical_potential}(b), and refer to this single value as $\mu(N)$. We summarize the values $\mu(N)$ determined this way for all relevant parameters to this manuscript in table~\ref{tbl:mu}

\begin{table}[]
\label{tbl:mu}
    \centering
\begin{tabular}{|c|c|c|}
\hline
n.n. hopping & hole-doping $p$ & $\mu(N)$ \\
\hline
$t'=-0.2$  & 1/16 &  1.733   \\
$t'=0$  & 0& 1.644 \\
$t'=0$  & 1/16& 1.602 \\
$t'=0.2$ & 1/16 & 1.695   \\
\hline
\end{tabular}
    \caption{Values of $\mu(N)$ for simulations parameters in the is manuscript on the $16\times 4$ cylinder wth $J/t=0.4$.}
    \label{tab:placeholder}
\end{table}

\section{Pure $t-J$ model}
\label{app:heatmaps}

\begin{figure*}[t]
    \centering
    \includegraphics[width=0.49\linewidth]{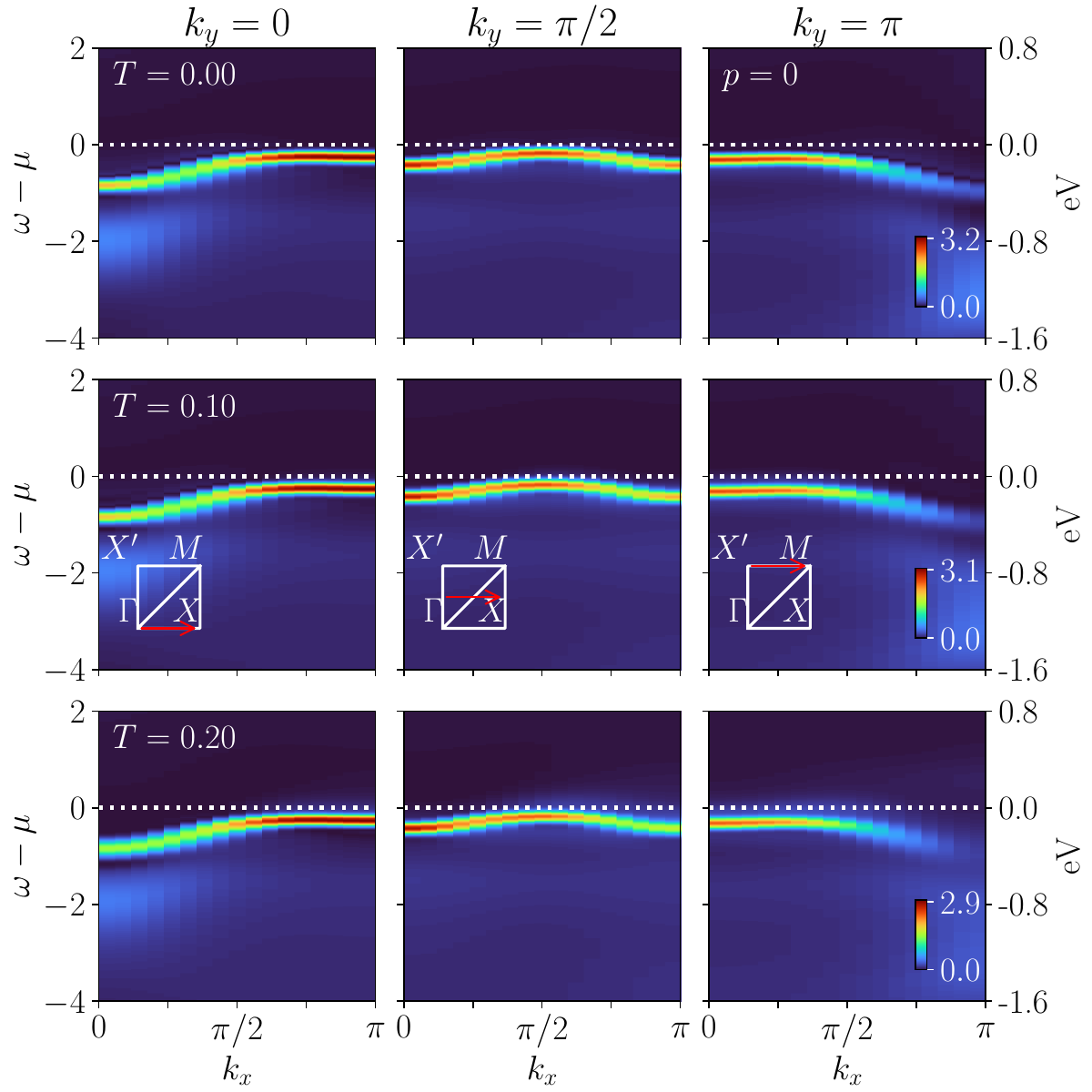}
    \includegraphics[width=0.49\linewidth]{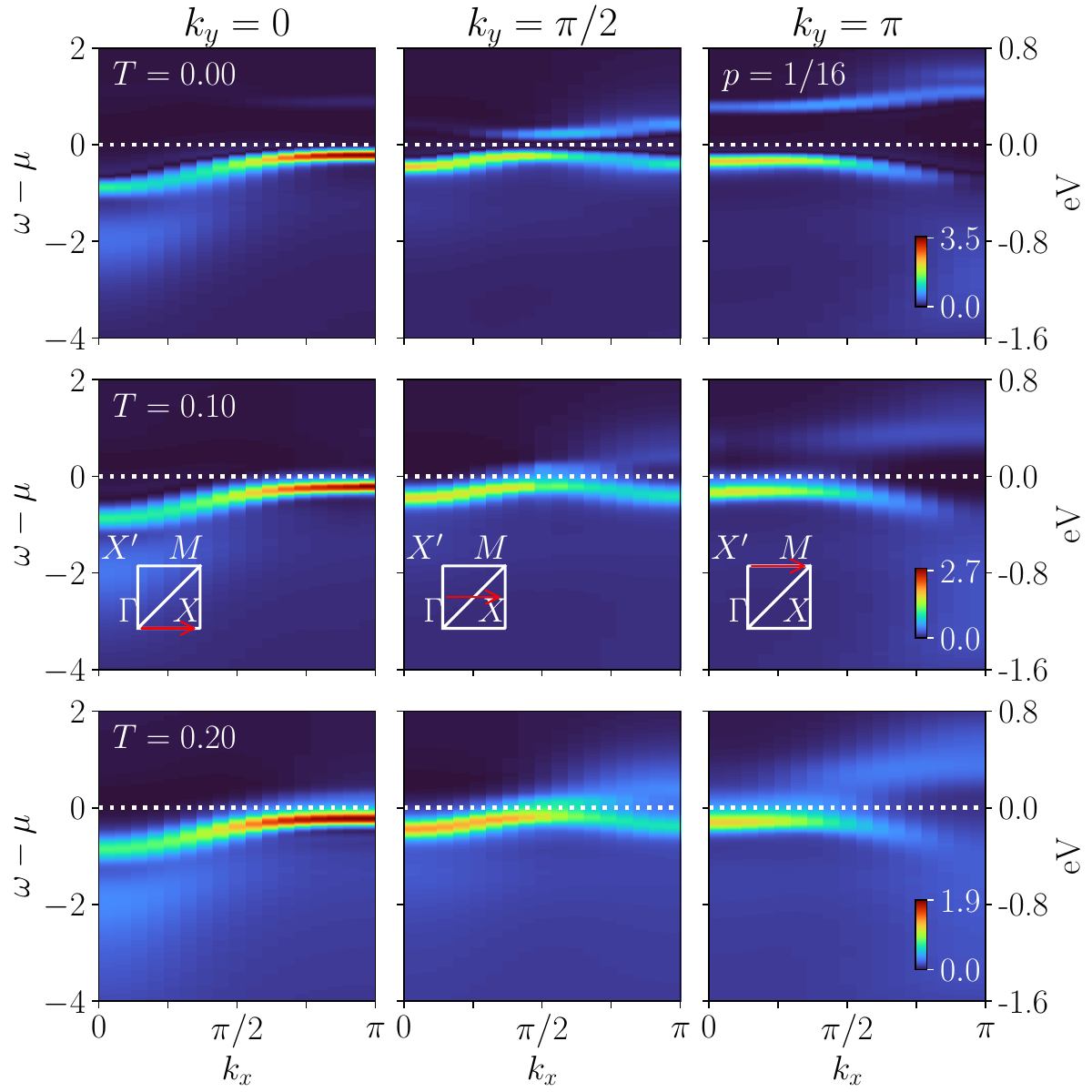}
    \caption{The spectral function $A({\bf k},\omega)$ in pure pure $t$-$J$ model ($t'=0$) for $p=0$ (left) and $p=1/16$ (right). For the time-evolution, $\theta=\pi/8$ is used.}
    \label{fig:pure_tj}
\end{figure*}

For reference, we computed the spectral function also in the pure ($t'=0$) $t$-$J$ model, which is presented in Fig.~\ref{fig:pure_tj} at two filling fractions. Since particle excitations are forbidden in the case $p=0$ by the Gutzwiller projection, only the hole spectral function appears (below the chemical potential). The appearance of the hole band in the undoped (Mott insulating) case allows us to establish the feature as evolving from a hole moving in the antiferromagnetic background. This band is subject to nontrivial renormalization upon introducing $t'$. 

\section{METTS and TDVP}
\label{app:tdvp}

\begin{figure}[ht!]
    \centering
    \includegraphics[width=0.99\linewidth]{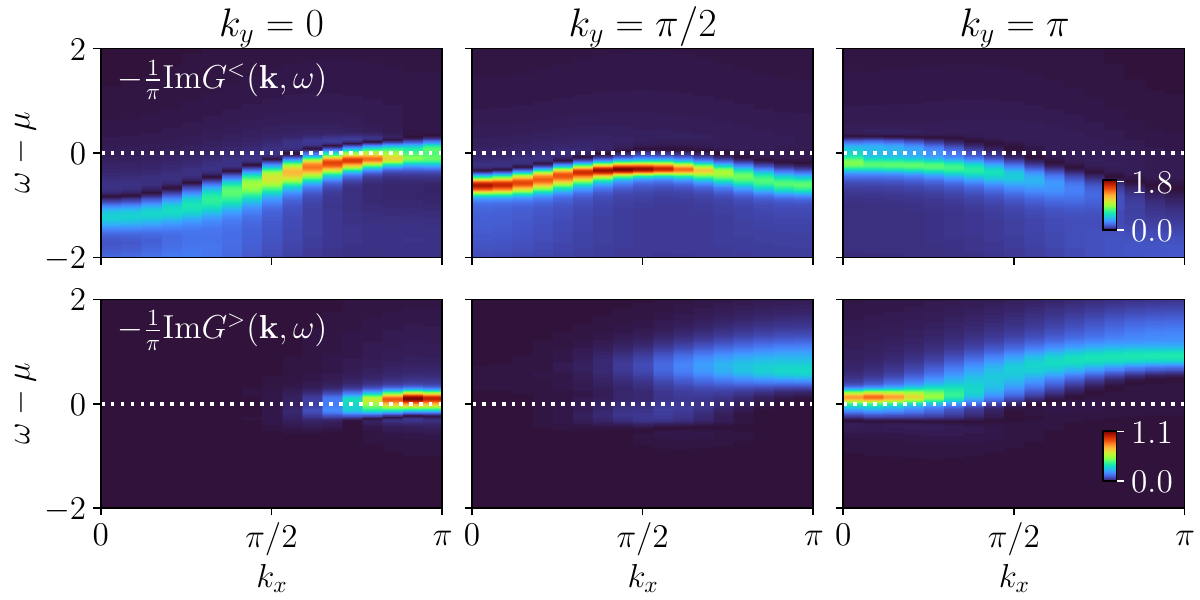}
    \caption{Breakdown of the spectral function into hole and particle contributions near the chemical potential. Shown are the imaginary parts of the lesser and greater Green's functions, Eq.~\ref{eq:lessr_greater_greens}. The parameters are $t'=0.2$, $p=1/16$, and $T=0.1$.}
    \label{fig:app_lg}
\end{figure}

Dynamical observables in the METTS ensemble are calculated as
\begin{align}
    \langle A(t)B\rangle_T&=Z^{-1}\sum_i p_i\langle\psi_i|A(t)B|\psi_i\rangle, \\ 
    |\psi_i\rangle&=\frac{e^{-\beta H/2}}{\sqrt{p_i}}|\sigma_i\rangle, \\ 
    p_i&=\langle\sigma_i|e^{-\beta H}|\sigma_i\rangle.
\end{align}
The quantities $p_i$ are interpreted as probabilities to sample the state $|\psi_i\rangle$ from the METTS ensemble at temperature $T=1/\beta$, $|\sigma_i\rangle$ are product states obtained by collapsing $|\psi_i\rangle$, and $Z=\sum_ip_i$ is the partition function. The spectral function is then $A_{\bf k}(\omega,T)=-\textrm{Im}G_{\bf k}(\omega,T)/\pi$. We use a version of the dynamical METTS algorithm~\cite{wang2025prb}, which requires the evaluation of two time-dependent states
\begin{align}
    |v_i(t)\rangle&=e^{-iHt}B|\psi_i\rangle,\\
    |w_i(t)\rangle&=e^{-iHt}|\psi_i\rangle,
\end{align}
\noindent
and their overlap
\begin{equation}
    C_i(t):=\langle w_i(t)|A|v_i(t)\rangle, \quad C_i(\omega)=\int dt e^{i\omega t} C_i(t).
\end{equation}

Since we are dealing with a cutoff in the maximum time $\Omega_t$ up to which we can perform the time evolution, we use a time-domain filter of the Hanning or Blackman-Harris form. Further, as the model is defined on a cylinder of finite length, we also employ a spatial filter, again of the Hanning or Blackman-Harris form, to alleviate the effects of the open boundary (which are observed to be quite minimal). Finally, we perform the sum over METTS samples $C(\omega)=\sum_i p_i C_i(\omega)$. Note that the (retarded) Green's function $G^{\textrm{R}}({\bf k},t)=\Theta(t)[G^>({\bf k},t)+\lbrace G^<({\bf k},t)\rbrace^*]$, where we introduced the greater and lesser Green's functions,
\begin{align}
    \nonumber
   G^<({\bf k},t)&=\ii\langle c^{\dagger}_{\bf k}(t)c_{\bf k}\rangle\\
   G^>({\bf k},t)&=-\ii\langle c_{\bf k}(t)c^{\dagger}_{\bf k}\rangle,
   \label{eq:lessr_greater_greens}
\end{align}
which are the actual quantities we compute. $G^<_{\bf k}$ contains information about hole excitations and reveals filled states, i.e., the ones visible to ARPES, while $G^>_{\bf k}$ sees particle excitations, such as seen in reverse photoemission experiments. The breakdown of the spectral function for the model considered here is illustrated in Fig.~\ref{fig:app_lg}

\begin{figure}[t]
    \centering
    \includegraphics[width=0.95\linewidth]{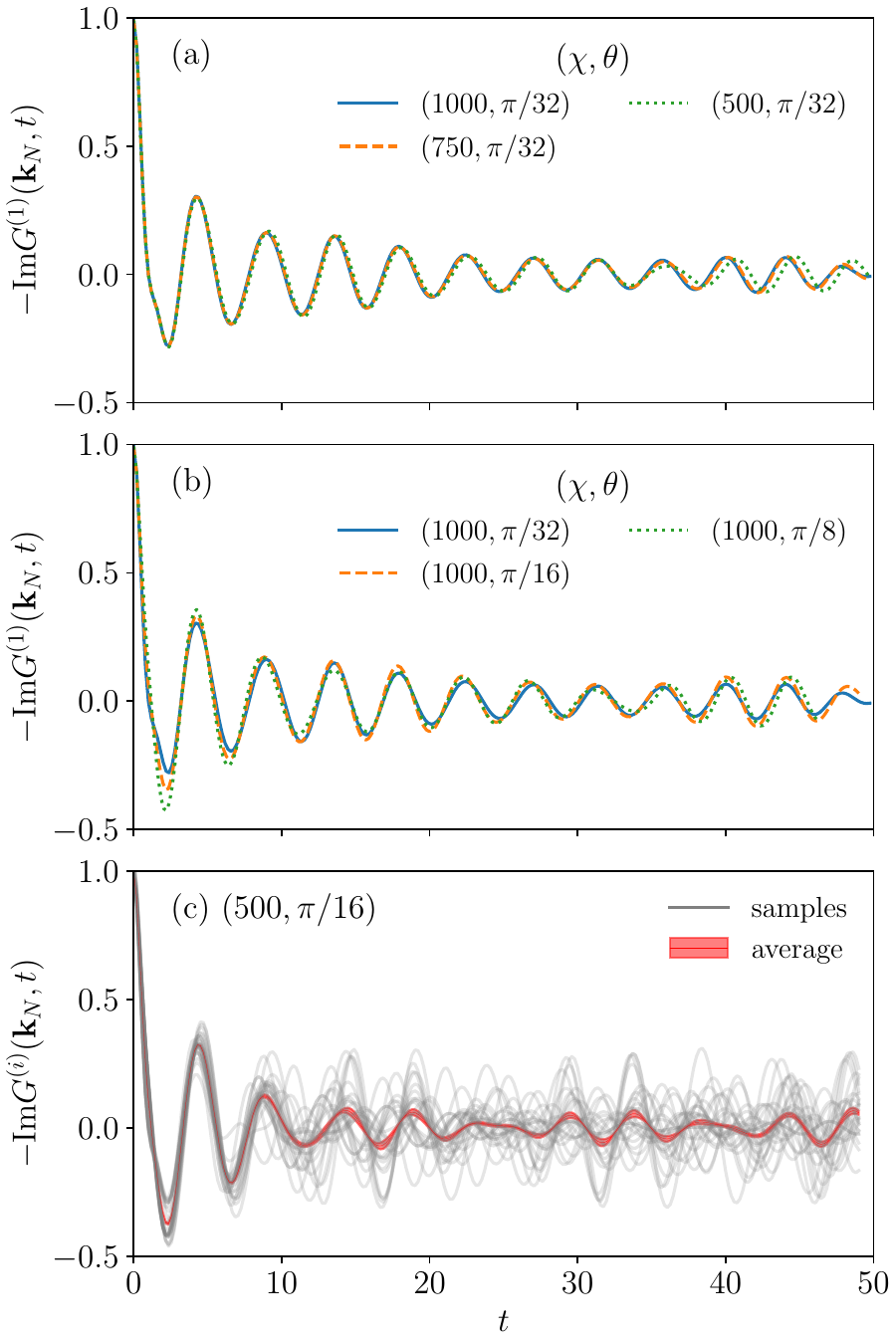}
    \caption{(a) The correlation function $C(t)$ from a single METTS snapshot, time-evolved with various $\chi$. (b) Same, but changing $\theta$. (c) $C_i(t)$ for several representative samples. No temporal filter is applied. In these cases, $T=0.1$, $t'=0$, and $p=1/16$.}
    \label{fig:s1}
\end{figure}

\begin{figure}[t]
    \centering
    \includegraphics[width=0.95\linewidth]{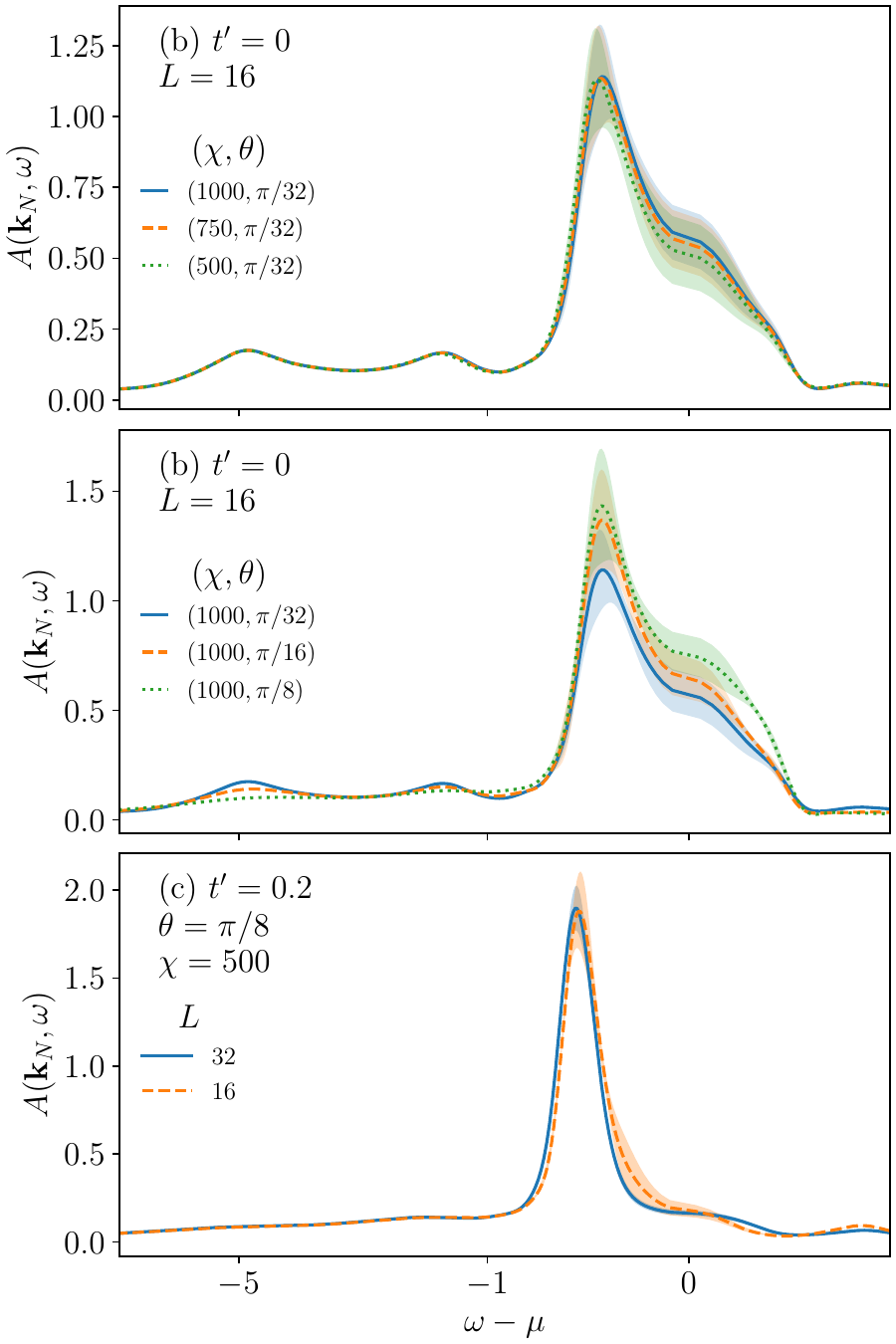}
    \caption{The spectral function $A({\bf k},\omega)$ at the node for several values of (a) bond dimension $\chi$ and (b) the angle $\theta$ in the complex plane. (c) The same quantity on cylinders of size $4\times L$. In both cases, $T=0.1$ and $p=1/16$. The chemical potential for the $L=32$ system is $\mu=1.703$. Note that the frequency axis is distorted as $\mathrm{sign}(\omega-\mu)\times (\omega-\mu)^2$ to facilitate comparison.
    }
    \label{fig:s2}
\end{figure}

To handle the time evolution, we use an algorithm based on the time-dependent variational principle (TDVP)~\cite{haegeman2011} with a complex time parameter~\cite{wang2025prb}, given by $U(\tau)=\exp(-\tau H)$, $\tau=t(i\cos\theta-\sin\theta)$ with $\theta\leq\pi/8$. The introduction of a complex time comes with the advantage that the entanglement content of the resulting time-dependent METTS is diminished, allowing one to reach longer times $\Omega_t$ with a modest bond dimension $\chi$. We fix the bond dimension $\chi_0$ during imaginary time evolution to produce METTS to $\chi_0=2000$. During the second (complex) time evolution, we vary the parameters $\chi$ and $\theta$ and present the analysis in Figs.~\ref{fig:s1} and~\ref{fig:s2}. We find good agreement between the various choices of $\chi=500-1000$, even for the smaller choice of $\theta$. Increasing $\theta$ has been observed to slightly dampen high-frequency contributions to spectral functions~\cite{wang2025prb}, which is, in our case, apparent in the incoherent continuum around $\omega-\mu\sim -5$, see Fig.~\ref{fig:s2}(b). We note that similar structures in the continuum have been observed previously~\cite{guthardt2025prb}. Finally, Fig.~\ref{fig:s2}(c) contains a study of the robustness of the obtained results for cylinders of different length $L$. We have also benchmarked the METTS-based approach against ED on a smaller system. 

The sampling requirements for convergence in principle depend on $T$, i.e., in general, more samples are required at higher $T$. To avoid autocorrelation between the samples in the METTS Markov chains, we consider every at most every 10th sample from each chain, with typically 10-40 chains per $T$. We, however, observe a significant degree of self-averaging in $G_i(t)$ with the biggest uncertainty stemming from the quasiparticle peak position.

\section{Clustering from METTS snapshots}
\label{app:clustering}
\begin{figure*}[t]
\centering
\includegraphics[width=0.95\linewidth]{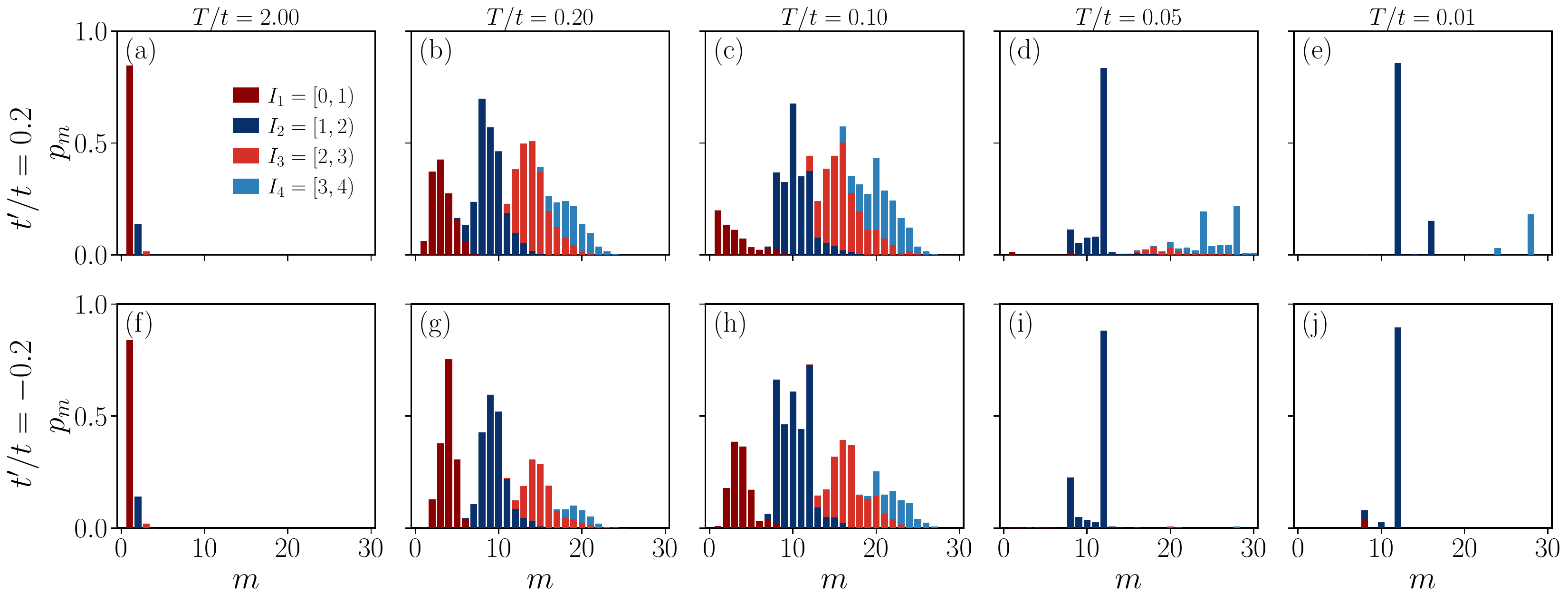}
\caption{Density-weighted cluster-size distributions $p_m$ for $p=1/16$ and $t'/t=\pm0.2$, resolved by cluster hole-mass bins $I_1=[0,1)$, $I_2=[1,2)$, $I_3=[2,3)$, and $I_4=[3,4)$. Columns correspond to $T/t=2.00,\,0.20,\,0.10,\,0.05,\,0.02$, and rows to $t'/t=0.2$ and $t'/t=-0.2$. The distribution $p_m$ is plotted against the cluster size $m$, the total bar height gives the fraction of doped charge residing in clusters of size $m$, and the stacked colors indicate how many holes those clusters carry. Upon cooling, the broad intermediate-temperature clustered regime reorganizes into a low-temperature CDW cluster.}
\label{fig:app_clustering}
\end{figure*}
To quantify real-space charge inhomogeneity in the finite-temperature METTS ensemble, we analyze the site-resolved hole density in individual snapshots. For a METTS snapshot $\ket{\psi_s}$, we define the local hole density
\begin{equation}
n_h^{(s)}(\mathbf r)=1-\langle \psi_s|n(\mathbf r)|\psi_s\rangle,
\label{eq:app_nh_snapshot}
\end{equation}
where $n(\mathbf r)=\sum_\sigma c^\dagger_{\mathbf r\sigma}c_{\mathbf r\sigma}$ is the local electron density operator. To identify hole-rich regions, we compute the spatial mean and variance within each snapshot,
\begin{equation}
\begin{aligned}
\overline{n}_h^{(s)} &= \frac{1}{N_s}\sum_{\mathbf r} n_h^{(s)}(\mathbf r), \\
\left(\sigma_{n_h}^{(s)}\right)^2 &= \frac{1}{N_s}
\sum_{\mathbf r}
\left(n_h^{(s)}(\mathbf r)-\overline{n}_h^{(s)}\right)^2,
\end{aligned}
\end{equation}
and introduce the adaptive threshold
\begin{equation}
n_{h,\rm th}^{(s)}=\overline{n}_h^{(s)}+c\,\sigma_{n_h}^{(s)},
\label{eq:app_threshold}
\end{equation}
with $c=0.5$. A site is classified as hole-rich if $n_h^{(s)}(\mathbf r)\ge n_{h,\rm th}^{(s)}$, i.e.,
\begin{equation}
\eta^{(s)}(\mathbf r)=
\begin{cases}
1, & n_h^{(s)}(\mathbf r)\ge n_{h,\rm th}^{(s)},\\
0, & \text{otherwise}.
\end{cases}
\end{equation}
Clusters are defined as connected components of the set $\{\mathbf r:\eta^{(s)}(\mathbf r)=1\}$ using nearest-neighbor connectivity on the cylindrical lattice (periodic in $y$, open in $x$). For each cluster $\mathcal C$, we record its size $m_{\mathcal C}=|\mathcal C|$ and its hole mass
$M_{\mathcal C}=\sum_{\mathbf r\in\mathcal C} n_h^{(s)}(\mathbf r)$. The main observable is the density-weighted cluster-size distribution,
\begin{equation}
p_m=
\frac{
\displaystyle
\sum_{s}\;\sum_{\mathcal C\in s:\,m_{\mathcal C}=m} M_{\mathcal C}
}{
\displaystyle
\sum_{s}\;\sum_{\mathcal C\in s} M_{\mathcal C}
},
\label{eq:app_pm_final}
\end{equation}
which measures the fraction of the total doped charge residing in clusters of size $m$ across the full METTS ensemble. By construction, $\sum_m p_m=1$. To obtain the stacked histograms, we resolve $p_m$ by cluster hole mass. We partition the mass into unit intervals
\begin{equation}
I_k=[k-1,k), \qquad k=1,2,3,\dots,
\end{equation}
and define
\begin{equation}
p_m^{(k)}=
\frac{
\displaystyle
\sum_{s}\;\sum_{\substack{\mathcal C\in s:\\ m_{\mathcal C}=m,\, M_{\mathcal C}\in I_k}} M_{\mathcal C}
}{
\displaystyle
\sum_{s}\;\sum_{\mathcal C\in s} M_{\mathcal C}
},
\label{pmk}
\end{equation}
such that
\begin{equation}
p_m=\sum_{k\ge1} p_m^{(k)}.
\end{equation}

\noindent The total bar height $p_m$ gives the fraction of doped charge residing in clusters of that size, while the color decomposition indicates how many holes those clusters carry. This provides a real-space characterization of how doped charge organizes across the METTS ensemble.
In Fig.~\ref{fig:app_clustering}, we show $p_{m}$ vs $m$ for cylinders of size $16\times 4$, doping $p=1/16$ for the $t'$-$t$-$J$ model both for $t'/t=0.2$ and $t'/t=-0.2$. We can identify three regimes. At high temperature, most of the weight is concentrated at the smallest cluster sizes, indicating weak real-space clustering of the holes. Upon cooling to $T/t\approx 0.1$-$0.2$, the distribution broadens strongly, with substantial weight over a wide range of $m$, signaling an intermediate regime of fluctuating mesoscopic charge clusters. At the lowest temperatures, the weight collapses onto a narrow peak at a characteristic size, corresponding to the CDW pattern selected by the finite cylinder geometry.

\begin{figure}[t]
    \centering
    \includegraphics[width=0.78\linewidth]{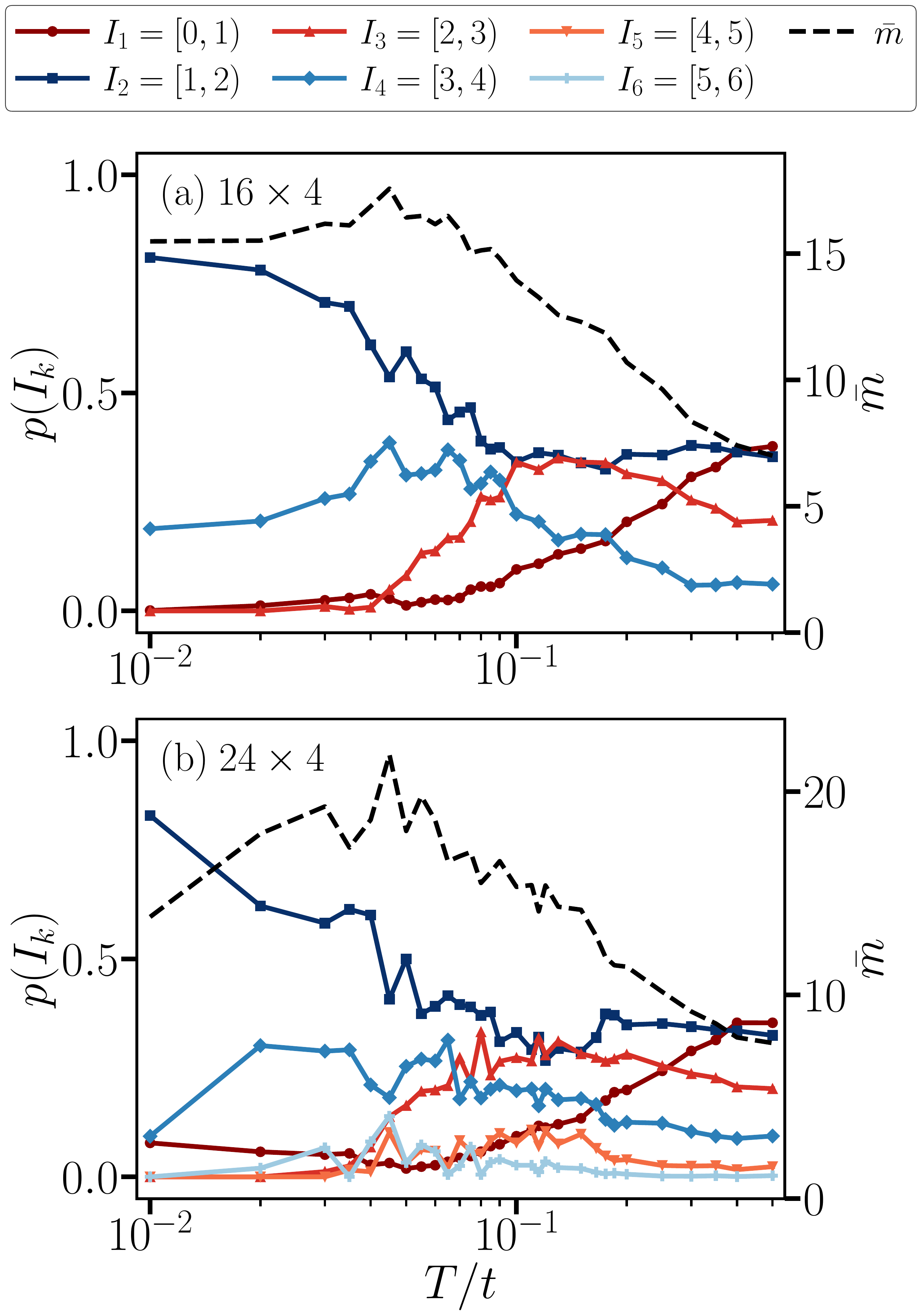}
\caption{Temperature evolution of the cluster hole-mass breakdown for $t'/t=0.2$
and doping $p=1/16$ on the (a) $16\times4$ and (b) $24\times4$ cylinders.
The left axis shows $p(I_k)=\sum_m p_m^{(k)}$, where $p_m^{(k)}$ is defined
in Eq.~\eqref{pmk}, and the dashed black curve shows the mean cluster size
$\bar m$ on the right axis. Only the physically available hole-mass sectors
are shown: $I_1$--$I_4$ for the $16\times4$ cylinder, which contains four
holes, and $I_1$--$I_6$ for the $24\times4$ cylinder, which contains six holes.
We see a transfer of weight from $I_1$ toward $I_2$ upon cooling. In both
panels, $I_3$ is finite at intermediate $T$.}
\label{fig:appendix_mass_sector_temperature}
\end{figure}

We also calculated the fraction of the total clustered hole mass carried by clusters whose hole mass lies in the interval
$I_k$,
\begin{equation}
p(I_k)\equiv \sum_m p_m^{(k)},
\end{equation}
with $p_m^{(k)}$ defined above in Eq.~\eqref{pmk}, and show it in Fig~\ref{fig:appendix_mass_sector_temperature}. The quantity measures how the clustered charge is partitioned among different hole-mass sectors.

For the $16\times4$ cylinder, Fig.~\ref{fig:appendix_mass_sector_temperature}(a), $I_1$ is the dominant sector at large $T$, but loses its weight upon cooling. Conversely, $I_2$ grows and becomes dominant in the CDW-ordered low-$T$ state. Importantly, the odd-hole number sector $I_3$ is also visible: it carries appreciable weight at intermediate temperatures, but is suppressed again at lower temperatures. The same tendencies are clearer for the $24\times4$ cylinder, see Fig.~\ref{fig:appendix_mass_sector_temperature}(b). The intermediate-temperature regime also contains visible contributions from $I_3$, and, more weakly, higher mass sectors (with the odd-hole $I_5$ present more strongly than the even $I_6$). The appearance of odd-hole clusters is significant because they cannot be the result of a paired state of two holes (heuristically, a ``pre-formed pair'') or combinations thereof.

\end{document}